\documentclass{webofc}
\usepackage[varg]{txfonts}   % Web of Conferences font
\begin{document}
\title{Modelling the cold dust in nearby spiral galaxies with radiative transfer}

\author{\firstname{Angelos} \lastname{Nersesian}\inst{1}\fnsep\thanks{\email{angelos.nersesian@ugent.be}} \and  
\firstname{Maarten} \lastname{Baes}\inst{1} \and 
\firstname{Suzanne C.} \lastname{Madden}\inst{2}}

\institute{Sterrenkundig Observatorium Universiteit Gent, Krijgslaan 281 S9, B-9000 Gent, Belgium 
\and Laboratoire AIM, CEA/DSM - CNRS - Universit\'{e} Paris Diderot, IRFU/Service d’Astrophysique, CEA Saclay, 91191, Gif-sur- Yvette, France}

\abstract{%
  Cosmic dust grains are one of the fundamental ingredients of the interstellar medium (ISM). In spite of their small contribution to the total mass budget, dust grains play a significant role in the physical and chemical evolution of galaxies. Over the past decades, a plethora of multi-wavelength data, from UV to far-infrared, has increased substantially our knowledge on the dust properties of nearby galaxies. Nevertheless, one regime of the spectrum, the mm range, remains relatively unexplored. Thanks to the new, high-resolution data in the mm range observed with the NIKA2 instrument and our radiative transfer framework, we aim to firmly characterise the physical properties of the very cold dust (<15K), and to quantify the importance of different emission mechanisms in the mm. So far, we have developed a methodology to use dust radiative transfer modelling and applied it to a small group of face-on spiral galaxies. The combination of the new NIKA2 data with our radiative transfer techniques would provide the right conditions to generate an accurate model of the interplay between starlight and dust in a sizeable sample of spatially-resolved nearby galaxies. 
}
\maketitle
\section{Introduction} \label{intro}

Galaxies are complex structures of multiple components: dark matter, stars, interstellar gas and dust. In particular, dust grains are carbonaceous or silicate based particles in the interstellar medium (ISM) of galaxies, and are responsible for the attenuation and reddening effects at ultraviolet (UV) and optical wavelengths. The role of dust does not stop there, since it is involved in several physical and chemical processes, by catalysing the formation of molecular hydrogen, and regulating the heating or cooling of the interstellar gas. One of the more pronounced characteristics of dust is that it absorbs roughly 30\% of the light emitted by stars and redistributes this light back in the mid- and far-infrared wavelengths \cite{Popescu_2002ApJ...567..221P, Skibba_2011ApJ...738...89S, Viaene_2016A&A...586A..13V, Bianchi_2018A&A...620A.112B}. The emission in that wavelength regime is often associated and used as a tracer of the star-formation activity in galaxies \cite{Calzetti_2007ApJ...666..870C, Calzetti_2010ApJ...714.1256C, Kennicutt_2009ApJ...703.1672K}. However, the contribution of stars of age greater than 8~Gyr to the dust heating can be non-negligible \cite[e.g.][]{Bendo_2012MNRAS.423..197B, Bendo_2015MNRAS.448..135B} and needs to be considered while estimating the rate which galaxies form their stars. 

One of the more advanced methods at our disposal to retrieve the distribution and physical properties of dust in galaxies, is by modelling the interplay between starlight and dust through radiative transfer (RT). RT modelling allows the generation of a realistic model of the radiation field and its interaction with dust, by quantifying the relative contribution of each stellar population to the dust heating. The advantage of using RT in relation to other methods, such as pixel-by-pixel SED (spectral energy distribution) fitting, is that it takes into account the effect of non-local heating in a 3D geometrical setting. In a recent series of radiative transfer studies, \cite{De_Looze_2014A&A...571A..69D} developed a method to quantify the dust heating fraction related to the star-formation activity in face-on spiral galaxies, by constructing the RT model of M~51a (Paper I). The authors showcased the importance of the older stellar population ($\sim$10~Gyr) to the heating of the diffuse dust in the ISM, donating up to 35\% of its total energy budget to that cause. Within the DustPedia\footnote{\url{http://dustpedia.astro.noa.gr}} framework \cite{Davies_2017PASP..129d4102D} the methodology of constructing an RT model was optimised and applied to the early-spiral galaxy M~81 \cite[Paper II,][]{Verstocken_2020A&A...637A..24V}, to a sample of face-on barred galaxies \cite[Paper III,][]{Nersesian_2020A&A...637A..25N}, to NGC~1068 which hosts an active galactic nucleus (AGN) \cite[Paper IV,][]{Viaene_2020A&A...638A.150V}, and to the M~51 interacting system \cite[Paper V,][]{Nersesian_2020A&A...643A..90N}. Two additional studies followed the same strategy of building models for two galaxy members of the Local Group, M~31 \cite{Viaene_2017A&A...599A..64V}, and M~33 \cite{Williams_2019MNRAS.487.2753W}.

In this proceedings paper, we would like to present few of the highlight-results of this series of radiative transfer papers, to inform about the limitations of our methodology, and how we can treat them using the newly available observational data in the mm regime from the NIKA2 instrument.  

\section{Modelling approach} \label{modelling_approach}

\begin{figure}[h]
\begin{center}
\includegraphics[scale=0.6]{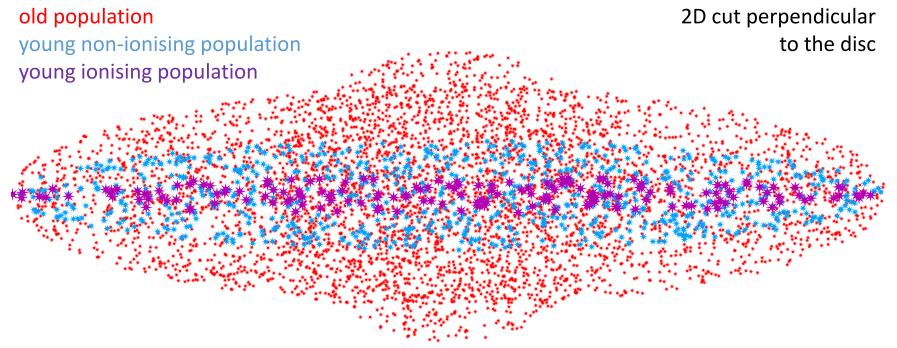}
\label{fig-1}       
\caption{A schematic of the standard geometric model of spiral galaxies adopted in our studies. The model consists of an old stellar bulge and disk component, and two young stellar populations: a non-ionising (100~Myr), and an ionising (10~Myr). The dust distribution is assumed to have the same scale-height as the young non-ionising stars. Image Credit: \citep{Verstocken_2020A&A...637A..24V}.}
\end{center}
\end{figure}

In order to generate a self-consistent 3D model for a specific galaxy, we use the state-of-the-art RT code \texttt{SKIRT} \cite{Baes_2011ApJS..196...22B, Camps_2015A&C.....9...20C, Camps_2020A&C....3100381C}. \texttt{SKIRT} embraces the Monte-Carlo method, taking into account all relevant physical processes such as scattering, absorption and thermal re-emission by dust. Figure~1 shows a schematic representation of the standard geometric model of spiral galaxies adopted in our work. The model consists of four stellar components and a dust component. We consider a central stellar bulge, and a disk populated by an old and young (non-ionising and ionising) stellar populations. We model the old and young non-ionising stellar populations using the Bruzual~\&~Charlot \cite{Bruzual_2003MNRAS.344.1000B} SSP of ages 8~Gyr and 100~Myr respectively, while for the young ionising stellar population we adopt the SED templates from MAPPINGS III \cite{Groves_2008ApJS..176..438G} assuming an age of 10~Myr. 

Broadband images at different wavelengths and at their native resolution were used to retrieve the geometric distributions of the various stellar and dust components. A brief description of the modelling steps follows. First, a bulge-to-disk decomposition is performed using an IRAC~3.6~$\mu$m image, tracing the emission of old stars. The bulge is approximated with a flattened S\'{e}rsic profile \cite{Sersic_1963BAAA....6...41S}. Then the old stellar population disk is obtained by subtracting the bulge from the total observed 3.6~$\mu$m emission. Similarly, we retrieve the distributions of the young non-ionising and ionising stars in the disk from the GALEX far-ultraviolet (FUV) image, and the combination of the MIPS 24~$\mu$m image with an H$\alpha$ map, respectively. The dust component is constrained through a FUV attenuation map, while for the dust composition we used the dust model \texttt{THEMIS} \cite{Jones_2017A&A...602A..46J}.

The 3D distribution of each of the disk components is introduced by assigning an exponential profile of different scale-heights based on previous estimates of the vertical extent of edge-on galaxies \cite{De_Geyter_2015MNRAS.451.1728D}. Based on the dust map a binary tree dust grid is generated \cite{Saftly_2014A&A...561A..77S}, through which photons propagate in our simulations. Each galaxy's dust grid has approximately 3 million dust cells. In general, we have three free parameters in our model determined via a $\chi^2$ optimisation procedure. These parameters are: the FUV luminosities of the young and ionising stellar populations and the dust mass. In special occasions, e.g. in the case of AGN emission, more free parameters are introduced in the model. Initial guesses for these parameters are based on global SED fitting results \cite{Nersesian_2019A&A...624A..80N}.

\section{Results} \label{results}

\begin{figure}[h]
\begin{center}
\includegraphics[scale=0.5]{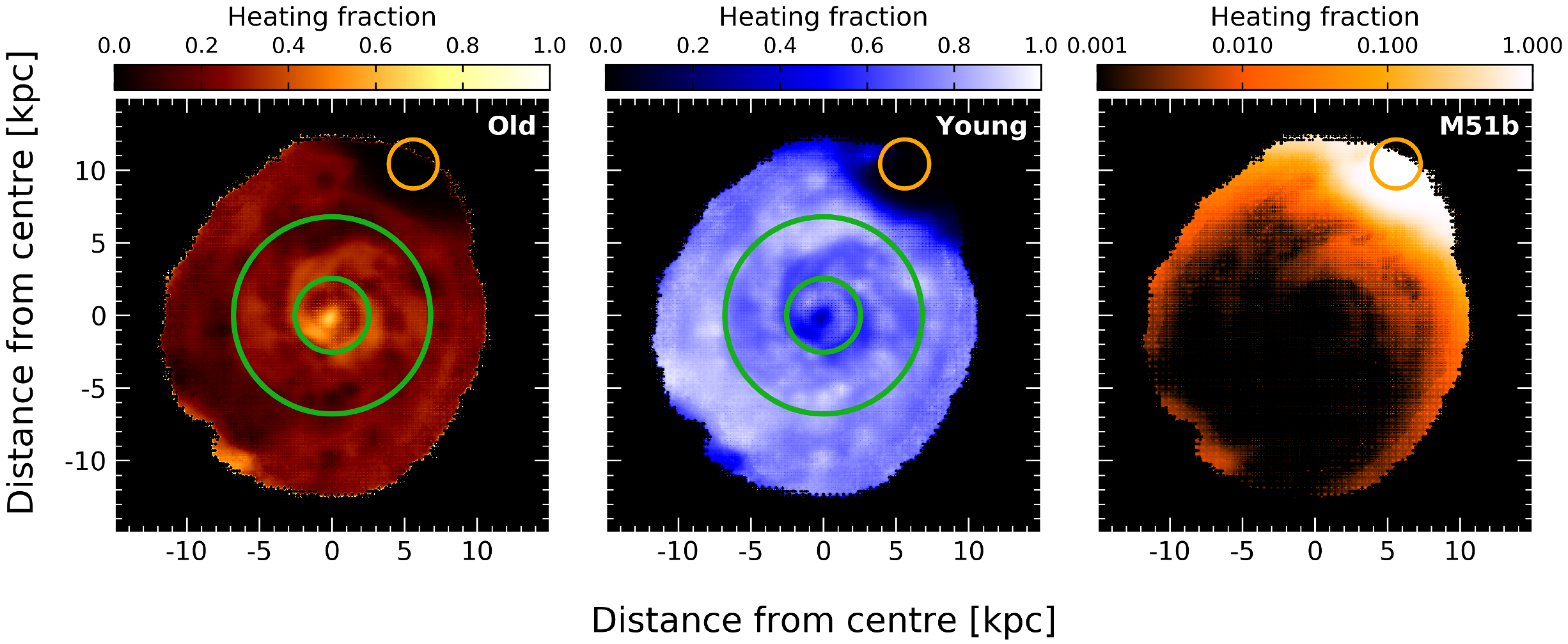}
\label{fig-2}       
\caption{Dust heating maps of a face-on view of M~51. Left panel: dust heating fraction by the old stellar population; middle panel: dust heating fraction by the young stellar populations; and right panel: dust heating fraction by M~51b. The gold circle indicates the extent of the dusty disk of M~51b. Image Credit: \cite{Nersesian_2020A&A...643A..90N}.}
\end{center}
\end{figure}

Throughout this series of radiative transfer papers new insights on the dust heating processes were presented. \cite{Verstocken_2020A&A...637A..24V} showed that the contribution of the old stellar population to the dust heating of M~81 is approximately 50\%, highlighting the importance of old stars to the FIR emission. Furthermore, \cite{Nersesian_2020A&A...637A..25N} studied a small sample of barred galaxies (M~83, M~95, M~100, and NGC~1365) and presented evidence of how the galactic bars may affect the processes of star formation and dust heating. The estimated contribution of the old stellar population in those galaxies remains considerably high, yet not as not as significant as in M~81. 

Viaene et al. \cite{Viaene_2020A&A...638A.150V} demonstrated that the AGN in NGC~1068 (M~77) is only important to dust heating within the inner few hundred parsecs from the centre of the galaxy, while most of the FIR emission is powered by star formation. The most recent RT study by \cite{Nersesian_2020A&A...643A..90N} investigated how the companion galaxy (M~51b) of the M~51 interacting system affects the SED, and consequently the dust heating of the main galaxy (M~51a). For that reason, a 3D RT model of the system was constructed. Figure~2 depicts the various dust-heating fractions from a face-on view of M~51. The left and middle panels show the dust-heating maps for the old and young stellar populations, respectively. The right panel depicts the dust-heating fraction provided by the older stellar population of the companion galaxy. The authors found that for the combined M~51 system, on average, 5.8\% of the dust emission is attributed to the dust heating by the old stellar population of M~51b, of which 4.8\% occurs in the M~51a subsystem; 23\% of the dust heating is supplied by the old stellar population of M~51a, while the remaining 71.2\% arises from the heating by the young stellar populations. 

From the 3D model of each galaxy we can immediately quantify the contribution of different sources (young stars, evolved stellar populations) to the dust. Our RT models suggest a strong correlation between the dust heating fraction and the specific star-formation rate (sSFR). In Fig.~3 we plot the heating fraction for six galaxies, as a function of the sSFR. It is immediately evident that there is an increasing trend between the heating fraction and the sSFR. This, in principle, enables us to quantify the heating fraction based on sSFR measurements in other galaxies. As we discuss later, our plan is to uniformly model a significant sample of galaxies which will allow us to confirm these trends and to firmly quantify this relation.

\begin{figure}[h]
\begin{center}
\includegraphics[scale=0.6]{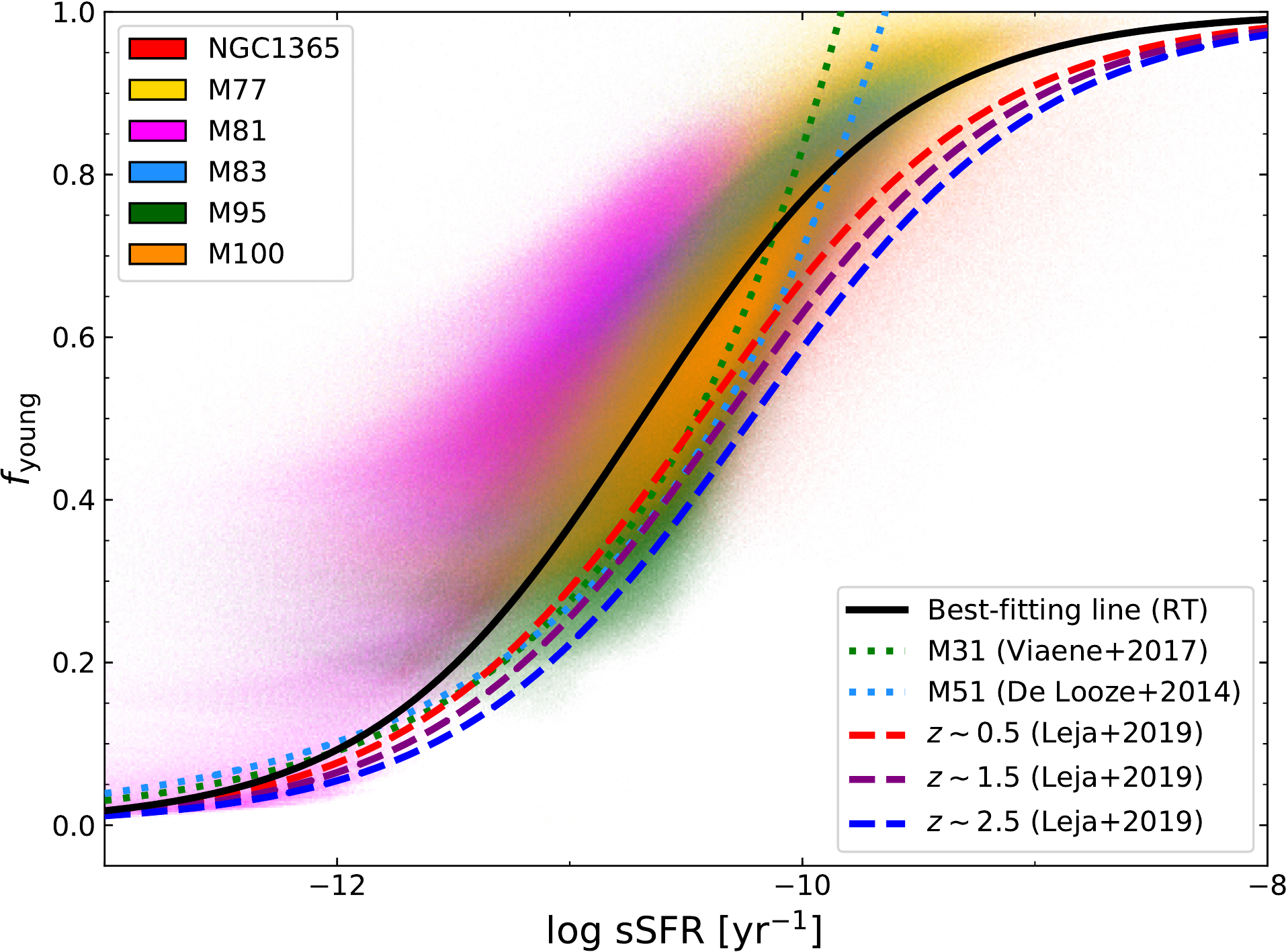}
\label{fig-3}       
\caption{The relation between the fraction of dust heating by young stars and the specific star formation rate of galaxies. Image Credit: \cite{Nersesian_2020A&A...637A..25N}.}
\end{center}
\end{figure}

In spite of its clear advancement compared to the previous generation of galaxy dust radiative
transfer models, this approach still has a number of persistent limitations. One of the most important limitations is the insensitivity of the DustPedia data set to cold dust. Although the \textit{Herschel} Space Observatory delivered new data that expanded our view on the dust properties, they are still limited to 500~$\mu$m which may not be sufficient to trace any cold dust ($<15~$K). Of course, \textit{Planck} data cover longer wavelengths, but unfortunately they lack the necessary spatial resolution. The lack of high-resolution observations at submm/mm wavelengths hinders our ability to firmly determine the dust emissivity index, and to trace the cold dust component assumed to carry most of the dust mass in galaxies \cite{Siebenmorgen_1999A&A...351..495S, Bendo_2006ApJ...652..283B}. Apart from cold dust emission, synchrotron emission (dominated by emission from supernova remnants) and free-free emission (from ionised gas in star-forming regions) are the two other dominant processes, which we need to quantitatively assess and disentangle from the dust emission.

Fortunately, this issue can now be addressed thanks to a new development. New observational imaging data on nearby galaxies are currently being taken in the critical submm/mm wavelength range. The most important data are coming from the NIKA2 camera (1.2 and 2~mm) on the IRAM 30~m telescope, and the SCUBA2 camera (850~$\mu$m) on the JCMT telescope. In particular, the IMEGIN project (PI: Suzanne Madden) is an ongoing 200~h program dedicated to deep NIKA2 continuum imaging of a set of 22 nearby spiral galaxies. All IMEGIN galaxies are taken from the DustPedia sample and hence have excellent ancillary UV-FIR data available, as well as complementary radio, molecular and atomic gas observations. Moreover, many of the IMEGIN galaxies are also targeted with SCUBA2 in the frame of the DOWSING survey, and plans are currently made to also observe the remaining galaxies with SCUBA2. Combining all these ingredients will allow us to move to a next level of 3D dust radiative transfer modelling of nearby galaxies, and to use these models to address a number of the fundamental questions on interstellar dust in galaxies.

\section{Conclusions} \label{conclusions}

We have constructed highly detailed 3D RT models for a small group of face-on galaxies to investigate the dust heating processes. We find that the contribution from the older stellar population to the dust emission at the infrared bands is more significant than previously thought. This result questions the use of infrared luminosity as a proxy for the star formation activity in star-forming galaxies. Furthermore, we confirm a tight correlation between the dust heating fraction and the sSFR. The full description of our methodology and the results of the modelling of M~81 are presented in \cite{Verstocken_2020A&A...637A..24V}, while the detailed results for the barred galaxies are discussed in \citep{Nersesian_2020A&A...637A..25N}. Finally, the modelling of a galaxy with an AGN component, NGC~1068 is presented in \citep{Viaene_2020A&A...638A.150V}, and the results of the M~51 interacting system in \citep{Nersesian_2020A&A...643A..90N}.

\section*{Acknowledgements}

AN, gratefully acknowledges the support of the Research Foundation - Flanders (FWO Vlaanderen). DustPedia is a collaborative focused research project supported by the European Union under the Seventh Framework Programme (2007-2013) call (proposal no. 606847). The participating institutions are: Cardiff University, UK; National Observatory of Athens, Greece; Ghent University, Belgium; Universit\'{e} Paris Sud, France; National Institute for Astrophysics, Italy and CEA, France. This work was supported by the Programme National "Physique et
Chimie du Milieu Interstellaire" (PCMI) of CNRS/INSU with INC/INP co-funded by CEA and CNES.


\begin{thebibliography}{}

\bibitem{Popescu_2002ApJ...567..221P}
C.C. Popescu, \textit{et al.}, ApJ \textbf{567}, 221 (2002)

\bibitem{Skibba_2011ApJ...738...89S}
R.A. Skibba, \textit{et al.}, ApJ \textbf{738}, 89 (2011)

\bibitem{Viaene_2016A&A...586A..13V}
S. Viaene, \textit{et al.}, A\&A \textbf{586}, A13 (2016)

\bibitem{Bianchi_2018A&A...620A.112B}
S. Bianchi, \textit{et al.}, A\&A \textbf{620}, A112 (2018)

\bibitem{Calzetti_2007ApJ...666..870C}
D. Calzetti, \textit{et al.}, ApJ \textbf{666}, 870 (2007) 

\bibitem{Calzetti_2010ApJ...714.1256C}
D. Calzetti, \textit{et al.}, ApJ \textbf{714}, 1256 (2010)

\bibitem{Kennicutt_2009ApJ...703.1672K}
R.C. Kennicutt, \textit{et al.}, ApJ \textbf{703}, 1672 (2009)

\bibitem{Bendo_2012MNRAS.423..197B}
G.J. Bendo, F. Galliano, S.C. Madden, MNRAS \textbf{423}, 197 (2012)

\bibitem{Bendo_2015MNRAS.448..135B}
G.J. Bendo, \textit{et al.}, MNRAS \textbf{448}, 135 (2015)

\bibitem{De_Looze_2014A&A...571A..69D}
I. De Looze, \textit{et al.}, A\&A \textbf{571}, A69 (2014) 

\bibitem{Davies_2017PASP..129d4102D}
J.J. Davies, \textit{et al.}, PASP \textbf{129}, 044102 (2017)

\bibitem{Verstocken_2020A&A...637A..24V}
S. Verstocken, \textit{et al.}, A\&A \textbf{637}, A24 (2020)

\bibitem{Nersesian_2020A&A...637A..25N}
A. Nersesian, \textit{et al.}, A\&A \textbf{637}, A25 (2020)

\bibitem{Viaene_2020A&A...638A.150V}
S. Viaene, \textit{et al.}, A\&A \textbf{638}, A150 (2020)

\bibitem{Nersesian_2020A&A...643A..90N}
A. Nersesian, \textit{et al.}, A\&A \textbf{643}, A90 (2020)

\bibitem{Viaene_2017A&A...599A..64V}
S. Viaene, \textit{et al.}, A\&A \textbf{599}, A64 (2017)

\bibitem{Williams_2019MNRAS.487.2753W}
T.G. Williams, \textit{et al.}, MNRAS \textbf{487}, 2753 (2019)

\bibitem{Baes_2011ApJS..196...22B}
M. Baes, \textit{et al.}, ApJS \textbf{196}, 22 (2011)

\bibitem{Camps_2015A&C.....9...20C}
P. Camps, M. Baes, Astronomy and Computing \textbf{9}, 20 (2015)

\bibitem{Camps_2020A&C....3100381C}
P. Camps, M. Baes, Astronomy and Computing \textbf{31}, 100381 (2020)

\bibitem{Bruzual_2003MNRAS.344.1000B}
G. Bruzual, S. Charlot, MNRAS \textbf{344}, 1000 (2003)

\bibitem{Groves_2008ApJS..176..438G}
B. Groves, \textit{et al.}, ApJS \textbf{176}, 438 (2008)

\bibitem{Sersic_1963BAAA....6...41S}
 J.L. S{\'e}rsic, Boletin de la Asociacion Argentina de Astronomia La Plata Argentina \textbf{6}, 41 (1963)

\bibitem{Jones_2017A&A...602A..46J}
A.P. Jones, \textit{et al.}, A\&A \textbf{602}, A46 (2017)

\bibitem{De_Geyter_2015MNRAS.451.1728D}
G. De Geyter, \textit{et al.}, MNRAS \textbf{451}, 1728 (2015)

\bibitem{Saftly_2014A&A...561A..77S}
W. Saftly, M. Baes, P. Camps, A\&A \textbf{561}, A77 (2014)

\bibitem{Nersesian_2019A&A...624A..80N}
A. Nersesian, \textit{et al.}, A\&A \textbf{624}

\bibitem{Siebenmorgen_1999A&A...351..495S}
R. Siebenmorgen, E. Kr{\"u}gel, R. Chini, A\&A \textbf{351}, 495 (1999)

\bibitem{Bendo_2006ApJ...652..283B}
G.J. Bendo, \textit{et al.}, ApJ \textbf{652}, 283 (2006)

\end{thebibliography}
\end{document}